\documentclass[aps,prl,twocolumn,showpacs,preprintnumbers]{revtex4}
\usepackage{textcomp}
\usepackage{amsmath}
\usepackage{amssymb}
\usepackage{graphicx}

\begin{document}

\title{Optical interface created by laser-cooled atoms trapped in the evanescent field surrounding an optical nanofiber}

\author{E.~Vetsch, D.~Reitz, G.~Sagu\'e, R.~Schmidt, S.~T.~Dawkins, and A.~Rauschenbeutel}

\email{rauschenbeutel@uni-mainz.de}

\affiliation{Institut f\"ur Physik, Johannes
Gutenberg-Universit\"at Mainz, 55099 Mainz}

\date{\today}

\begin{abstract}

Trapping and optically interfacing laser-cooled neutral atoms is
an essential requirement for their use in advanced quantum
technologies. Here we simultaneously realize both of these tasks
with cesium atoms interacting with a multi-color evanescent field
surrounding an optical nanofiber. The atoms are localized in a
one-dimensional optical lattice about 200~nm above the nanofiber
surface and can be efficiently interrogated with a resonant light
field sent through the nanofiber. Our technique opens the route
towards the direct integration of laser-cooled atomic ensembles
within fiber networks, an important prerequisite for large scale
quantum communication schemes. Moreover, it is ideally suited to
the realization of hybrid quantum systems that combine atoms with,
e.g., solid state quantum devices.

\end{abstract}


\maketitle

Laser-trapped atoms are well isolated from their environment and
can be coherently manipulated as well as efficiently interrogated
using resonant light \cite{Metcalf}. This makes them prime
candidates for the implementation of quantum memories and quantum
repeaters, necessary, e.g., for the operation of long distance
quantum communication networks \cite{BDCZ,Duan,Yuan,Zhao}. At the
same time, solid state quantum devices, such as quantum dots or
superconducting circuits, are readily miniaturized and integrated
using well established technologies \cite{Schoel}. For these
reasons, the possibility of combining atomic and solid state
devices in so-called hybrid quantum systems, which combine the
advantageous properties of both approaches, has recently attracted
considerable interest \cite{Andre,Monroe,Chang}. Two prerequisites
have to be fulfilled in order to realize such a hybrid quantum
system. On the one hand, the atoms would have to be efficiently
interfaced with resonant probe light for manipulation and
interrogation. On the other hand, the atoms would have to be
trapped in order to be placed in close vicinity of the charged or
magnetized solid state devices in order to be coupled via electric
or magnetic interaction. Here, we demonstrate that both tasks,
trapping and optically interfacing neutral atoms, can be
simultaneously accomplished by means of tapered optical fibers
with a nanofiber waist.

\begin{figure}[htp]
 \centering
  \includegraphics[width=0.41\textwidth]{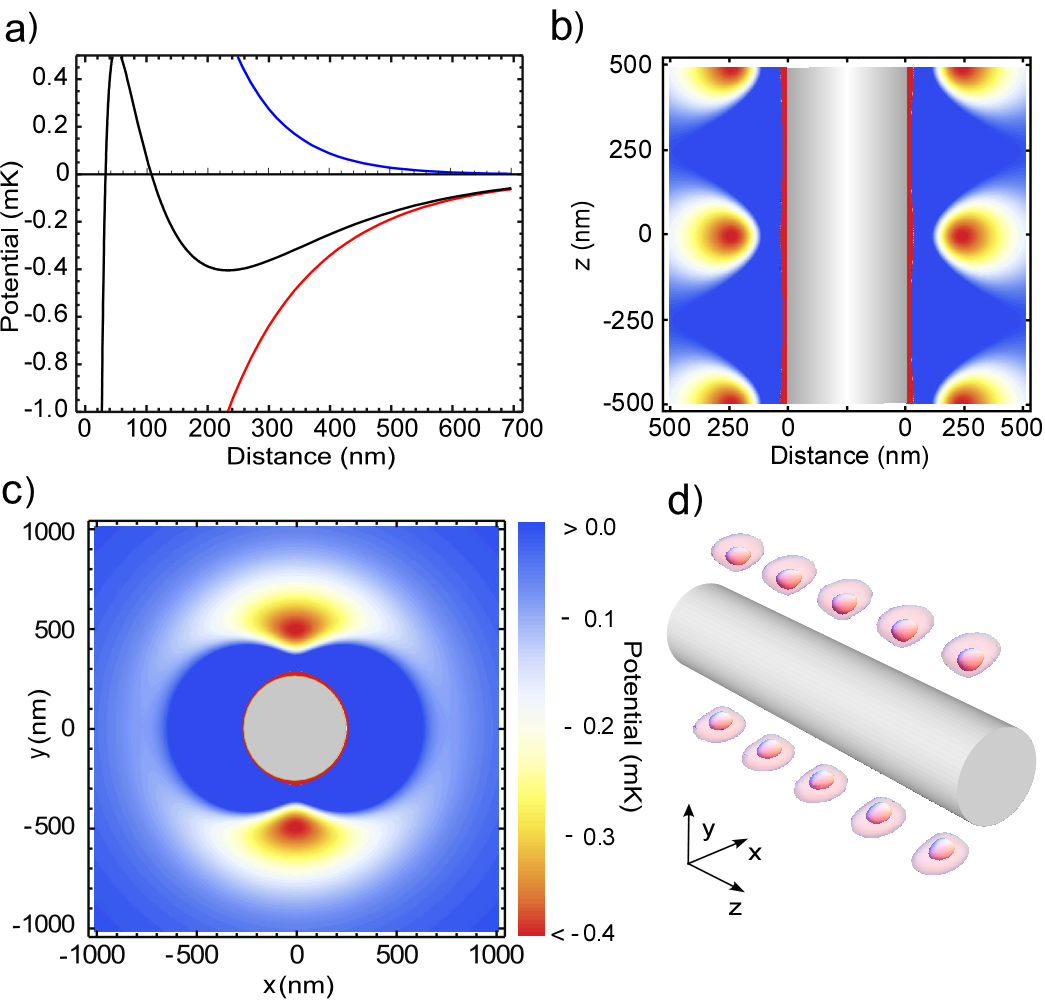}
 \caption{\label{fig:potential}
(a, black line) Potential as a function of distance from the
surface of a 500-nm diameter nanofiber for a ground state cesium
atom induced by a two-color evanescent field plus the van der
Waals potential. The red and blue lines show the individual light
induced potentials for a red- and blue-detuned field at 1064~nm
and 780~nm respectively. We assumed a power of $P_{\rm red} = 2
\times 2.2~$~mW (standing wave) and $P_{\rm blue} = 25$~mW and
orthogonal linear polarization of the red and blue fields. (b)
Contour plots of the same potential as in (a). The red-detuned
standing wave ensures axial confinement. (c) Azimuthal plot of the
same potential as in (a) and (b). The planes of the plots in
(a)--(c) are chosen to include the trapping minima. (d) Contour
plot of the resulting array of trapping sites on both sides of the
fiber showing equipotential surfaces 40~$\mu$K and 125~$\mu$K
above the trapping minimum.}
\end{figure}

The coupling of laser cooled atoms with light by means of optical
fibers has been an active field of research over the past years.
For this purpose, two types of optical fibers have been employed:
In hollow core fibers, the atoms are funneled into a capillary in
the centre of the fiber where they couple to the guided fiber mode
\cite{Chris,Bajcsy}. This implies that at least one end of the
hollow core fiber has to terminate inside the vacuum chamber and
that the device can thus not be directly connected to a fiber
network. The situation is different when using optical nanofibers
with a diameter smaller than the wavelength of the guided light. In
this case, the atoms remain at the outside of a fiber and couple to
the evanescent field surrounding the fiber \cite{Sague,Nayak}. Such
nanofibers can be realized as the waist of tapered optical fibers
(TOFs) which allows one to optimally match the mode of a standard
single mode optical fiber with the fundamental nanofiber mode, thus
ensuring high transmission of the device \cite{Love,Birks} as well
as direct integrability into fiber networks. Moreover, a nanofiber
can be passed through an operating magneto-optical trap (MOT),
thereby facilitating the coupling of atoms and light \cite{Sague,
Morrissey}. The ultimate goal in both lines of research is to
combine the coupling scheme with three dimensional trapping of the
atoms in order to maximize both the number of coupled atoms,
resulting in the highest possible optical depth, as well as the
interaction time. In this context, it has been proposed to realize
a two-color optical dipole trap which makes use of a red- and
blue-detuned evanescent light field around the optical nanofiber
\cite{Dowling,LeKien04}.

Figure~\ref{fig:potential} shows the potential calculated for a
ground state cesium atom subjected to such a two-color evanescent
light field around a 500-nm diameter nanofiber. At that diameter
the nanofiber only guides the fundamental HE$_{11}$ mode for both
wavelengths. The red-detuned light field, as well as the van der
Waals force attract the atoms towards the nanofiber while the
blue-detuned light field repels the atoms from the fiber. Due to
the different radial decay lengths of the red- and blue-detuned
evanescent fields one can thus create a radial potential minimum
at a few hundred nanometers from the nanofiber surface by properly
choosing the respective powers, see Fig.~\ref{fig:potential}~(a).
Confinement along the fiber axis is achieved by launching an
additional, counter-propagating red-detuned laser beam through the
fiber, thereby realizing a red-detuned standing wave, see
Fig.~\ref{fig:potential}~(b). Azimuthal confinement of the atoms
stems from the azimuthal dependence of the evanescent field
intensity for the quasi-linearly polarized HE$_{11}$ mode, see
Fig.~\ref{fig:potential}~(c). In order to maximize the azimuthal
confinement we use orthogonal linear polarizations for the red-
and blue-detuned fields. Figure~\ref{fig:potential}~(b) is shown
in the plane of polarization of the red-detuned field.
Figure~\ref{fig:potential}~(d) shows the two resulting 1d arrays
of trapping minima on both sides of the fiber, visualized by the
equipotential surfaces $40~\mu$K and $125~\mu$K above the trapping
minimum. The calculated trapping frequencies for the parameters
used in Fig.~\ref{fig:potential} are $200$~kHz, $315$~kHz, and
$140$~kHz in the radial, axial, and azimuthal direction,
respectively. Large detunings of the trapping light fields are
chosen in order to ensure a low scattering rate, compatible with a
theoretical coherence time of 50~ms and trap lifetime of up to
100~s. Further decoherence mechanisms due to surface interactions
are expected to be negligible at the distances of $\approx 100$~nm
considered here because of the low conductivity of glass
\cite{Henkel}.

\begin{figure}[htp]
 \centering
  \includegraphics[width=0.4\textwidth]{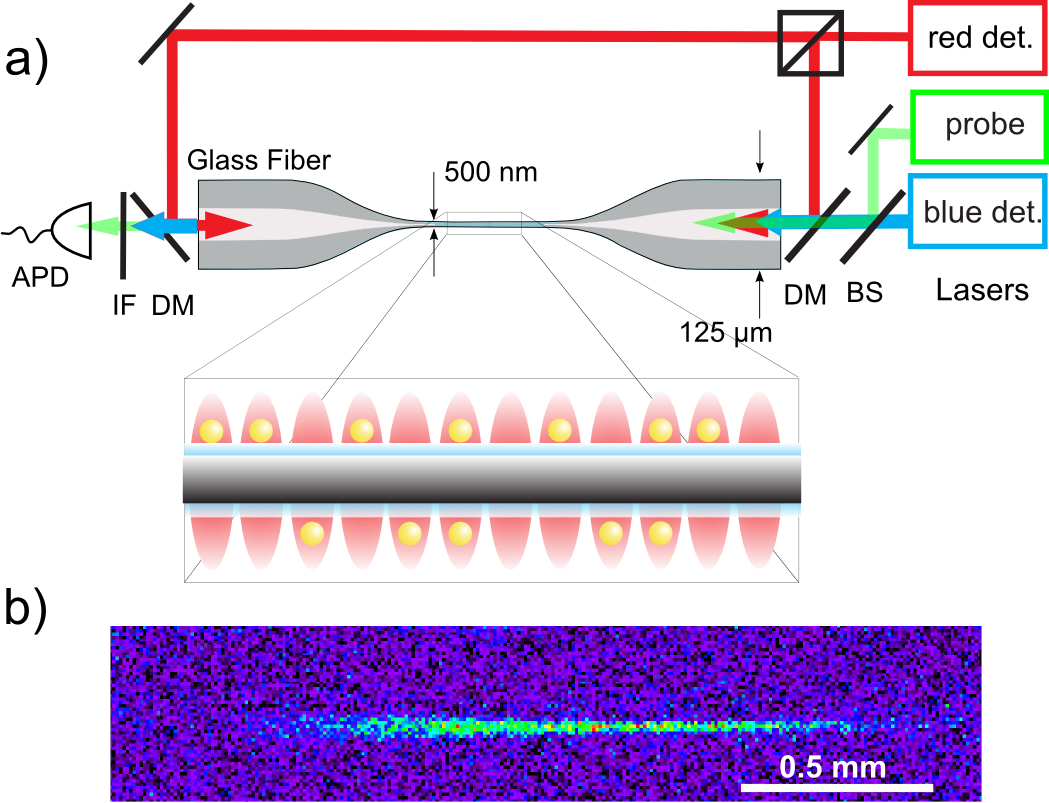}
  \caption{ \label{fig:setup}
(a) Experimental setup of the fiber-based atom trap. The
blue-detuned running wave in combination with the red-detuned
standing wave create the trapping potential. A resonant laser is
used for probing the atoms via the evanescent field. (b)
Fluorescence image of the trapped atomic ensemble.
  }
\end{figure}


Figure~\ref{fig:setup}~(a) shows a schematic of our setup. We use
a laser at 1064 nm which is red-detuned with respect to the D1
(894~nm) and D2 (852~nm) transitions of cesium. Using a beam
splitter, two beams are generated and coupled into both ends of
the TOF in order to realize the standing wave. A laser at 780 nm
is used for forming the blue-detuned potential and is superposed
with one of the red-detuned laser beams using a dichroic mirror
(DM). The probe laser is resonant with the D2 transition. It is
superposed with the blue-detuned laser beam using a beam splitter
(BS) and its power transmitted through the TOF ($\approx 1$~pW) is
measured using an avalanche photodiode (APD) in combination with
an interference filter (IF). Typical powers of the trapping lasers
are $2$--$4$~mW for each of the red-detuned beams and
$10$--$30$~mW for the blue-detuned beam. The state of polarization
of the dipole laser beams at the nanofiber waist can be monitored
and aligned  by observing the angular distribution of the Rayleigh
scattered light, emitted by the nanofiber perpendicular to its
axis using CCD cameras. The probe laser has a short term linewidth
of about 1~MHz.

The TOF has been fabricated by stretching a standard single mode
fiber while heating it with a travelling hydrogen/oxygen flame in
a computer controlled fiber pulling rig \cite{Warken}. The
nanofiber waist has a homogeneous diameter of 500 nm over its
length of 5~mm. In the 4~cm long tapered sections the weakly
guided LP$_{01}$ mode of the unstretched fiber is adiabatically
transformed into the strongly guided HE$_{11}$ mode of the
nanofiber waist and back \cite{Love,Birks}. This results in a
highly efficient coupling of light into and out of the nanofiber
yielding an overall transmission through the TOF of $97~\%$. The
lasers are coupled into the ends of the TOF using conventional
fiber couplers and the TOF enters and exits an ultra-high vacuum
chamber via a vacuum feed-through. Inside the chamber (pressure
$\approx 8\times 10^{-10}$~mbar), a six-beam magneto-optical trap
(MOT) produces a cold cesium atom cloud with a $1/e^{2}$-diameter
of $1.2$~mm which is spatially overlapped with the nanofiber.


Loading of the trap is accomplished as follows: The red- and
blue-detuned trapping light fields are present in the nanofiber at
all times. During the first 2~s, the atoms are captured and cooled
in the MOT. In the following $100$~ms the atoms are transferred
and cooled into the trapping minima along the nanofiber. For this
purpose, the power of the MOT cooling laser, repump laser and the
magnetic field gradient are ramped down to zero and the detuning
of the cooling laser is increased to about $-80$~MHz. We note that
due to the small trapping volumes the loading is expected to
operate in the so-called collisional blockade regime resulting in
an occupancy of at most one atom per trapping site
\cite{Schlosser}. The resulting maximum average occupancy of 0.5
in conjunction with the distance of $\approx 500$~nm between the
standing wave antinodes thus limits the maximum number of trapped
atoms to 2000 per millimeter.


In order to image the trapped atoms, we have positioned a
home-built microscope objective with a numerical aperture of
$0.29$ inside the vacuum chamber \cite{Alt}. This allows us to
collect the fluorescence light, induced by the resonant excitation
of the trapped atoms, and thus create an image on the chip of an
EMCCD camera. Figure~\ref{fig:setup}~(b) shows the fluorescence of
a trapped ensemble of Cs atoms recorded $20$~ms after loading the
trap. Fluorescence is observed over a length of $\approx 1.2$~mm
corresponding roughly to the diameter of the cold atom cloud in
the MOT. For this image the probe laser was detuned by $-10$~MHz
with respect to the AC-Stark shifted D2 ($F=4$ to $F'=5$)
transition of Cs and its power was set to $500$~pW, corresponding
to the saturation intensity at resonance at the position of the
atoms. The probe light was pulsed with a pulse length of 2~ms,
corresponding to the exposure time of the EMCCD camera. The
fluorescence image shown in Fig.~\ref{fig:setup}~(b) is the sum of
320 single background corrected exposures from consecutive
experimental runs.

\begin{figure}
 \centering
 \includegraphics[width=0.4\textwidth]{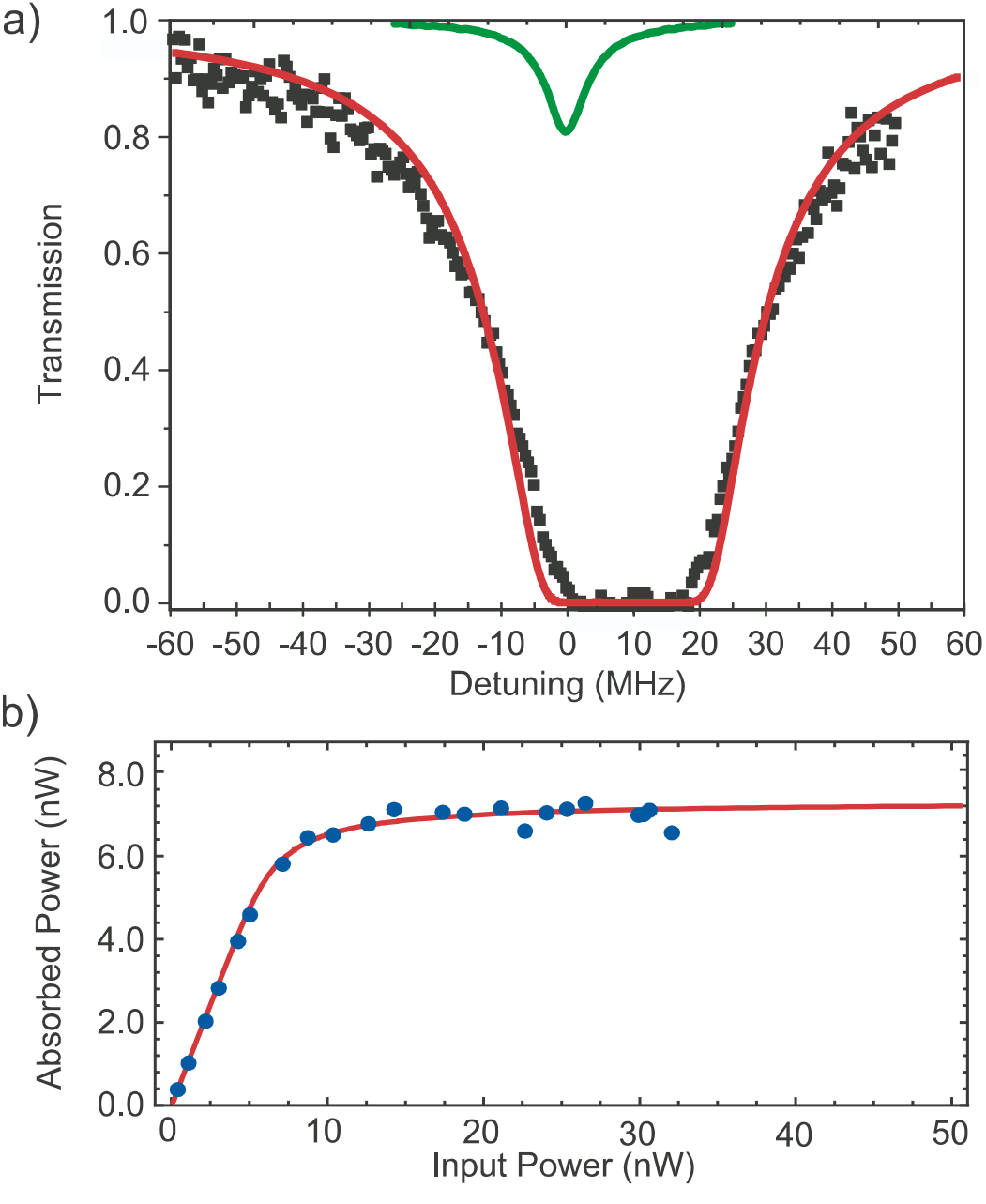}
  \caption{\label{fig:spectra}
(a) Transmission spectrum of the probe beam through the nanofiber
after loading the trap (black squares). For reference the spectrum
of the MOT-cloud (green line) is plotted. The red line is a
theoretical fit, see text. (b) Saturation measurement (blue
circles) yielding the number of trapped atoms. The red line is a
theoretical fit, see text. }
\end{figure}


We investigate the spectral properties of the trapped atomic
ensemble by measuring the absorption of the probe light as a
function of its detuning. As a reference, the green line in
Fig.~\ref{fig:spectra}~(a) shows the transmission of the probe
light versus the detuning with respect to the D2 ($F=4$ to $F'=5$)
transition of Cs without the fiber trap after abruptly switching
off the MOT lasers and magnetic field. In this case, the measured
absorption due to the cold atom cloud around the nanofiber reaches
$20$~\% and the linewidth is only slightly larger than the natural
linewidth of Cs due to the atom-surface interactions \cite{Sague}.
The black squares in Fig.~\ref{fig:spectra}~(a) show the probe
transmission directly after loading the fiber trap. We observe a
strong absorption resulting from a dramatic increase of the number
of atoms in the evanescent field due to trapping of atoms inside
the two-color trap.

The fitted line profile (solid red line) yields a maximum optical
depth of $OD = 13~(2)$ for a detuning of 13~MHz and a FWHM of
$\Gamma= 20$~MHz. The shift and broadening can be attributed to
the state dependent light-shift of the transition frequency
induced by the trapping laser fields. Compared to this
inhomogeneous broadening due to the transitions to all excited
substates the broadening due to the thermal motion of the atoms
inside the trapping potential is negligible. For the measurement
in Fig.~\ref{fig:spectra}~(a) the probe light power was about
1~pW. This low power ensures that the scattering rate of the atoms
(30~kHz) is smaller than their oscillation frequency in the trap
($\approx 140$~kHz), resulting in strongly suppressed recoil
heating due to off-resonant raman scattering \cite{Wolf}. This
maximizes the number of scattered photons before atom loss,
thereby optimizing the signal.

The APD signal is recorded with a digital storage oscilloscope and
averaged over 64 traces. The transmission is then calculated
according to
\begin{equation}
T = \frac{(P_{\rm at} - P_{\rm bg})}{(P_{0} - P_{\rm bg})},
\end{equation}
where $P_{\rm at}$, $P_{0}$, and $P_{\rm bg}$ are the APD signals
with and without atoms and the background signal, respectively.
The transmission spectrum in Fig.~\ref{fig:spectra}~(a) is well
described by
\begin{equation}
T(\Delta)=\exp\left\{-OD\sum_{i}\frac{C_{i}}{1+4(\Delta_{i}/\Gamma_{0})^{2}}\right\},
\end{equation}
(solid red line), where the exponent accounts for the Lorentzian
line profiles corresponding to the transitions between the
differently light-shifted new eigenstates,
$\Delta=\omega-\omega_{\rm D2}$ is the detuning of the probe laser
frequency with respect to the atomic resonance frequency in free
space $\omega_{\rm D2}$, and $\Delta_{i}=\Delta-\Delta_{i}^{\rm
LS}$, where $\Delta_{i}^{\rm LS}$ is the state dependent light
shift induced by the linearly polarized trapping lasers.
$\Gamma_{0}=5.2$~MHz is the natural line width of the Cs D2
transition. The coefficients $C_{i}$ account for the degeneracy
and relative strength of the transitions and are chosen such that
the sum in the exponent is normalized to one. Optical pumping
induced by the probe light as well as collective radiative effects
are not included in this model and might account for the slight
discrepancy between the theoretical prediction and the
experimental data.

In order to determine the number of trapped atoms, we carry out a
saturation measurement: We tune the probe laser to the Stark
shifted resonance of the trapped atoms and measure the absorbed
power as a function of the incident power, see
Fig.~\ref{fig:spectra}~(b). At high saturation, the atomic
ensemble absorbs about $P_{\rm Abs}\approx 7.5$~nW of probe light
power. By comparing this value with the power radiated by a single
fully saturated Cs atom of $P_{\rm Cs} = 3.8$~pW we infer that $N
= P_{\rm Abs}/P_{\rm Cs}= 2000$ atoms are present in the
fiber-based trap. The solid red line is a theoretical fit based on
a saturation model taking into account the spatially varying
intensity along the atomic ensemble due to absorption. We find
that the number of stored atoms decreases exponentially with a
time constant of about $50$~ms, smaller than what would be
expected from losses due to background gas collisions. This effect
is currently still under investigation, one possible loss
mechanism being heating of the atoms due to intensity fluctuations
of the trapping lasers.

From the optical depth and the number of atoms we infer an average
absorbance per atom of $\varepsilon=OD/N\approx 0.65$~\%. We note
that in the absence of inhomogeneous broadening the absorbance of
the atomic ensemble would be increased by a factor of $\approx
2.5$. Under these circumstances and for low saturation, the
absorbance per atom would then reach
$\varepsilon_{0}=\eta\varepsilon\approx 1.8$~\%. This value is
consistent with our expectation that a single atom at a radial
distance of about $230$~nm from the fiber surface would absorb a
fraction of  $\sigma/A_{\rm eff} = 1.5$~\%, where $\sigma$ and
$A_{\rm eff}$ are the absorption cross section of the unperturbed
D2 ($F=4$ to $F'=5$) transition of cesium and the effective mode
area of the nanofiber guided mode \cite{Kien06,Warken07}
respectively.

Summarizing, we have trapped and interfaced neutral cesium atoms
in a one-dimensional optical lattice created by a two-color
evanescent field surrounding an optical nanofiber. Due to the
atoms' very close proximity ($230$~nm) to the nanofiber surface,
the atoms efficiently interface with resonant light sent through
the nanofiber. We have observed an absorbance of about $0.7$~\%
per atom and an overall optical depth of $13$ for approximately
$2000$ trapped atoms. Our system is well suited to the realization
of hybrid quantum systems, coupling, e.g., solid state quantum
devices with optically interfaced laser-trapped atoms. Such a
coupling scheme would in particular profit from the possibility of
positioning the atoms in close vicinity of the solid state surface
without exposing the latter to the evanescent trapping light
field. Moreover, our work paves the way towards non-linear optics
and quantum communication applications with fiber-coupled atomic
ensembles. Finally, it should be possible to realize a
fiber-mediated coupling between the trapped atoms. In conjunction
with the trapping in 1d periodic chains, this would then allow the
study of collective states of light and matter, e.g., tailoring or
even fully suppressing the spontaneous emission of the ensemble
into free space \cite{Zoubi}.
\\
Acknowledgements: This work was supported by the Volkswagen
Foundation and the European Science Foundation. The authors wish
to thank Helmut Ritsch for fruitful discussions and helpful
suggestions, R. Mitsch for help with the data analysis and
LOT-Oriel for the loan of the EMCCD-camera.

\end{document}